\documentclass{PoS}

\title{Gauge mediation with light stops}

\ShortTitle{GMSB \& light stops}

\author{Antonio~Delgado~$^a$, Mateo~Garcia-Pepin~$^{bc}$ and \speaker{Mariano~Quiros}~$^{bcd}$\\
\llap{$^a$}  Department of Physics, 225 Nieuwland Science Hall, University of Notre Dame\\
Notre Dame, IN 46556, USA  \\
        \llap{$^b$} Institut de Fisica d'Altes Energies (IFAE),
        The Barcelona Institute of Science and Technology\\
        Campus UAB, 08193 Bellaterra (Barcelona) Spain\\
        \llap{$^c$} ICTP-SAIFR and Instituto de Fisica Teorica (IFT), Universidade Estadual Paulista (UNESP)\\
Rua Dr.  Bento Teobaldo Ferraz 271, 01140-070 Sao Paulo, SP Brazil \\
       \llap{$^d$} Institucio Catalana de Recerca i Estudis Avan\c{c}ats (ICREA)\\
        Campus UAB, 08193 Bellaterra (Barcelona) Spain\\
        E-mail: \email{antonio.delgado@nd.edu}, \email{mgarcia@ifae.es}, \email{quiros@ifae.es}}

\abstract{The mechanism of gauge mediated supersymmetry breaking (GMSB) solves the supersymmetric flavor problem although it requires superheavy stops to reproduce the experimental value ($125$~GeV) of the Higgs mass.  A possible way out is to extend the MSSM Higgs sector with triplets which provide extra tree-level corrections to the Higgs mass. Triplets with neutral components getting vacuum expectation values (VEV) have the problem of generating a tree-level correction to the $\rho$ parameter. We introduce supersymmetric triplets with hypercharges $Y=(0,\pm 1)$, with a tree-level custodial $SU(2)_L\otimes SU(2)_R$ global symmetry in the Higgs sector protecting the $\rho$ parameter: a supersymmetric generalization of the Georgi-Machacek model. The renormalization group running from the messenger to the electroweak scale mildly breaks the custodial symmetry. We will present realistic low-scale scenarios, their main features being a Bino-like neutralino or right-handed stau as the NLSP, light ($1$ TeV) stops, exotic couplings ($H^\pm W^\mp Z$ and $H^{\pm\pm} W^\mp W^\mp$) absent in the MSSM and proportional to the triplet VEV, and a possible (measurable)  universality breaking of the Higgs couplings $\lambda_{WZ}= \left(g_{hWW}/g_{hWW}^{SM}\right)/ \left(g_{hZZ}/g_{hZZ}^{SM}\right)\neq 1$.}

\FullConference{18th International Conference From the Planck Scale to the Electroweak Scale \\
		25-29 May 2015\\
		Ioannina, Greece }

\begin{document}

\section{Introduction}

ATLAS and CMS have discovered a scalar boson with properties consistent with those of the Standard Model (SM) Higgs and a mass $\sim 125$ GeV~\cite{Aad:2014aba,Khachatryan:2014jba}. Whether it actually is the SM depends on possible (future) deviations from the SM predictions in Higgs strengths (e.g.~in the $\gamma\gamma$, or any other channel).

However the SM suffers (as an effective theory with cutoff $\Lambda\sim M_P$) from a naturalness problem by which the Higgs mass receives huge (quadratic) corrections~\cite{Veltman:1980mj}
\begin{equation}
  \Delta m^2\simeq -\frac{3}{32\pi^2v^2}\left(m_H^2+2m_W^2+m_Z^2-4m_t^2  \right)\Lambda^2
  \end{equation}
 The paradigmatic solution to the problem of quadratic divergences is supersymmetry by which the previous correction cancels out. In particular the minimal SM supersymmetric extension (MSSM)~\footnote{For a recent analysis of quadratic divergences in the MSSM see Ref.~\cite{Masina:2015ixa} and references therein.}.

The mechanism of supersymmetry breaking is a big \textbf{unknown}. The standard assumption is that supersymmetry is broken in a hidden sector and transmitted to the observable sector either through gravitational, or gauge messengers. On the other hand supersymmetry can create an additional \textit{supersymmetric flavor problem} unless the messenger interaction is flavor diagonal. This is the case when messengers transmit supersymmetry breaking by gauge interactions. The so-called Gauge Mediated Supersymmetry Breaking (GMSB). In particular GMSB provides
\begin{itemize}
\item
Sfermion squared masses $m_{\tilde f}^2$ by two-loop diagrams
\item
Gaugino Majorana masses $M_a$ by one-loop diagrams
\item
Very small stop mixing $A_t$ by two-loop diagrams.
\end{itemize}
Because of the latter GMSB has a difficulty to reproduce large values of the Higgs mass and requires superheavy stop ($\tilde t$) masses. This problem has been analyzed in Refs~\cite{Draper:2011aa,Ajaib:2012vc}. The conclusion reached by these papers is that, for large values of $\tan\beta$, there is a lower bound on the stop mass as $m_{\tilde t}\gtrsim 10$ TeV. Of course for small values of $\tan\beta$, $\tan\beta\simeq 1$, the bound can be as much large as $m_{\tilde t}\sim 10^{10}$ GeV~\cite{Delgado:2013gza}.

Two options do appear to tackle this problem in GMSB theories without extending the SM gauge group:
\begin{enumerate}
\item
To introduce an \textit{extended} GMSB which generates large values of $A_t$ by direct messenger-MSSM superpotential couplings. However, extended GMSB does not necessarily lead to minimal flavor violation as the flavor constraints require a special flavor texture~\cite{Evans:2015swa}.
\item
To extend the MSSM adding extra tree-level contributions to the Higgs mass: Introducing singlets $S$ and/or $Y=(0,\pm 1)$, $\Sigma_{0,\pm1}$ triplets. Singlets do not acquire a mass unless in an extended GMSB scenario and triplets generate $\Delta\rho$ when getting a VEV $v_\Delta$.
\end{enumerate}
To tackle the $\rho$ problem in the presence of triplets  $\Sigma_{0,\pm1}$, either
$v_\Delta\to 0$ ($m^2_\Sigma\to\infty$) which is not possible in GMSB, or we
provide custodial global $SU(2)_L\otimes SU(2)_R$ symmetry to the triplet sector, which is the approach we will deal with in the present work. This approach was proposed in 1985 by Georgi and Machachek~\cite{Georgi:1985nv} in a seminal paper were they showed how to introduce triplet VEVs  while protecting the $\rho$ parameter by a custodial symmetry. This theory has been recently supersymmetrized in a series of papers, Refs.~\cite{Cort:2013foa,Garcia-Pepin:2014yfa,Delgado:2015aha,Delgado:2015bwa}, where a custodial symmetry is introduced in the Higgs superpotential and where the $\rho$ parameter is protected. This theory is dubbed supersymmetric custodial triplet model (SCTM) and will be used in this work.

\section{Overview of the SCTM}
We assume the supersymmetric theory to be invariant under $SU(2)_L\otimes SU(2)_R$, only  broken by Yukawa and hypercharge interactions. We add to the MSSM Higgs sector $H_1=(H_1^0,H_1^-)^T$ and $H_2=(H_2^+,H_2^0)^T$, with respective hypercharges $Y=(-1/2,\,1/2)$, three $SU(2)_L$ triplets $\Sigma_Y$ with hypercharges  $Y=(-1,\, 0,\, 1)$, which we represent by two dimensional matrices as~\cite{Cort:2013foa}
 \begin{equation}
 \Sigma_{-1}=\left(\begin{array}{cc} \frac{\chi^-}{\sqrt{2}} & \chi^0\\\chi^{--}& -\frac{\chi^-}{\sqrt{2}}
 \end{array}
 \right),\quad  \Sigma_{0}=\left(\begin{array}{cc} \frac{\phi^0}{\sqrt{2}} & \phi^+\\ \phi^{-}& -\frac{\phi^0}{\sqrt{2}}
 \end{array}
 \right),\quad  \Sigma_{1}=\left(\begin{array}{cc} \frac{\psi^+}{\sqrt{2}} & \psi^{++}\\\psi^{0}& -\frac{\psi^+}{\sqrt{2}}
 \end{array}
 \right)\ .
 \end{equation}
where $Q=T_{3L}+Y$. They are organized under $SU(2)_L\otimes SU(2)_R$ as $\bar H=(\textbf{2},\bar {\textbf{2}})$, and $\bar\Delta=(\textbf{3},\bar{\textbf{3}})$ where
\begin{equation}
\bar H=\left( \begin{array}{c}H_1\\ H_2\end{array}\right),\quad
\bar \Delta=\left(\begin{array}{cc} -\frac{\Sigma_0}{\sqrt{2}} & -\Sigma_{-1}\\ -\Sigma_{1}& \frac{\Sigma_0}{\sqrt{2}}\end{array}\right)
\end{equation}
and $\bar T_{3R}=-T_{3R}=Y$.   The invariant products for doublets $A\cdot B\equiv A^a\epsilon_{ab}B^b$  and anti-doublets $\bar A\cdot \bar B\equiv\bar A_a\epsilon^{ab}\bar B_c$ are defined by $\epsilon_{21}=\epsilon^{12}=1$. 

The total superpotential can be written as $W=W_{0}+W_Y$, where 
\begin{equation}
W_0=\lambda \bar H\cdot \bar\Delta\bar H+\frac{\lambda_3}{3}\textrm{tr} \bar\Delta^3+\frac{\mu}{2}\bar H\cdot\bar H+\frac{\mu_\Delta}{2}\textrm{tr} \bar\Delta^2 
\end{equation}
is the $SU(2)_L\otimes SU(2)_R$ invariant superpotential, while the Yukawa coupling superpotential  is defined as $W_Y=h_t\,\overline Q_L\cdot H_2 t_R + h_b\,\overline Q_L\cdot H_1 b_R+\cdots$.
Thus the pure Higgs sector superpotential respects the $SU(2)_L\otimes SU(2)_R$ invariance, while the superpotential Yukawa terms (as well as gauge terms provided by $U(1)_Y$ gauge interactions) explicitly break it. 

Gauge mediation will generate masses at the messenger scale $\mathcal M$.
The generated mass spectrum at $\mathcal M$ will be $SU(2)_L\otimes SU(2)_R$ invariant, 
except for the gauge contributions $\mathcal O(\alpha_1^2)$. However, this violation is similar to the violation of the custodial symmetry induced by the hypercharge coupling in the renormalization group (RG) running and does not spoil the main phenomenological features of the model.
The RG running will split the custodial invariants of the superpotential. The most general superpotential can then be written as
\begin{eqnarray}
W &= &- \lambda_aH_1\cdot\Sigma_1H_1+\lambda_b H_2\cdot\Sigma_{-1}H_2+\sqrt{2}\lambda_cH_1\cdot\Sigma_0H_2+\sqrt{2}\lambda_3\textrm{tr}\,\Sigma_1\Sigma_0\Sigma_{-1} \nonumber\\ & 
 - & \mu H_1\cdot H_2 + \frac{\mu_{a}}{2}\textrm{tr}\,\Sigma_0^2 + \mu_{b}\textrm{tr}\,\Sigma_1\Sigma_{-1}+ h_t\,\overline Q_L\cdot H_2 t_R + h_b\,\overline Q_L\cdot H_1 b_R
\label{superpotencial}
\end{eqnarray}
where the $SU(2)_L\otimes SU(2)_R$ invariant situation is recovered when $\lambda_a=\lambda_b=\lambda_c \equiv \lambda$ and $\mu_{a}=\mu_{b}\equiv \mu_\Delta$.

The total potential is then
$V=V_F+V_D+V_{\mathrm{soft}}$, where $V_F$ is the supersymmetric potential obtained from the superpotential (\ref{superpotencial}), $V_D$ is the potential from $D$-terms given in (\ref{potencialD}) and  $V_{\mathrm{soft}}$ the soft breaking potential given in Eq.~(\ref{potencialsoft}).
%
%
The $SU(2)_L\otimes SU(2)_R$ conditions in the supersymmetry breaking sector would be given by: $m_{H_1}=m_{H_2}\equiv m_H$, $m_{\Sigma_0}=m_{\Sigma_1}=m_{\Sigma_{-1}}\equiv m_{\Sigma}$, $B_{a}=B_{b}\equiv B_{\Delta}$ and $A_{a}=A_{b}=A_{c}\equiv A_\lambda$.

We now expand the neutral components of the fields in a totally general way as in Ref.~\cite{Garcia-Pepin:2014yfa}
$X =\frac{1}{\sqrt{2}}\left( v_X + X_R +\imath X_I \right)$, where $X=H^0_1,H^0_2,\phi^0,\chi^0,\psi^0$,
and we parametrize the departure from custodial symmetry through three angles as 
\begin{eqnarray}
v_1&=& \sqrt{2}\cos\beta v_H,~~v_2=\sqrt{2}\sin\beta v_H, \nonumber\\
v_\psi &=& 2\cos\theta_1\cos\theta_0 v_\Delta,~~v_\chi=2\sin\theta_1\cos\theta_0 v_\Delta,\nonumber\\
v_\phi &=& \sqrt{2} \sin\theta_0 v_\Delta .
\label{eq:vacio}
\end{eqnarray}
The parametrization preserves the relation $v^2 \equiv  2v_H^2+8v_{\Delta}^2$, and we recover the $SU(2)_V$ invariant vacuum when $\tan{\beta}=\tan{\theta_0}=\tan{\theta_1}=1$, $v_1=v_2\equiv v_H$ and $v_\psi=v_\chi=v_\phi\equiv v_\Delta $.

We now need to solve the equations of motion (EoM) ensuring correct EW breaking. Five neutral scalar fields will generate five minimization conditions that will fix five parameters. As the parameters $m_3^2$ and $B_{\Delta_{a,b}}$ have their RG equations decoupled from the rest, we can consistently fix two parameters, e.g.~$m_3^2$ and $B_{\Delta_a}$, from their respective EoMs. 
 The other three EoM self consistently determine the values of the custodial breaking angles ($\tan{\beta},\tan{\theta_0},\tan{\theta_1}$) which, given a set of custodial boundary conditions at the messenger scale, are a prediction of the EoMs for a given value of $v_\Delta$~\footnote{More details can be found in the Appendix.}. In turn this determines the $\rho$ parameter as $\rho=1+\Delta\rho$ where
\begin{equation}
\Delta\rho=\frac{2( 2v_\phi^2-v_\psi^2-v_\chi^2)}{v_1^2+v_2^2+4(v_\chi^2+v_\psi^2)} =-4\frac{\cos 2\theta_0 v_\Delta^2}{v_H^2+8\cos^2\theta_0v_\Delta^2}\ .
\end{equation}
Notice that only $\tan\theta_0$ affects the $\rho$ parameter. 

The EoMs are just criticality conditions as they do not tell us whether we are really exploring a minimum of the potential, and much less if this minimum is the absolute one. The minimum condition will be provided by the absence of tachyonic states in the scalar spectrum. Moreover each minimum we find is likely the deepest one since it consists on a smooth deformation of an $SU(2)_V$ preserving minimum where the D-terms vanish, therefore with minimized energy.

In the next section we will present a particular mechanism of gauge mediation by which soft masses are generated at the messenger scale $\mathcal M$ and the physical spectrum is obtained after running the soft masses down to the electroweak scale. As minimal gauge mediation (MGM)~\cite{Giudice:1998bp} provides a very rigid framework to encompass low energy phenomenology we will consider a particular model of general gauge mediation~\cite{Meade:2008wd} (GGM) where there is more flexibility to accommodate the supersymmetric mass spectrum of the SCTM. We will also consider low scale gauge mediation $\mathcal M=100$ TeV to minimize the effect of the custodially breaking running. As the group transmitting supersymmetry breaking is the SM group, the boundary conditions at $\mathcal M$ are custodial invariant except for the contribution of the $U(1)_Y$ which breaks as usual custodial symmetry.

\section{Gauge mediation in SCTM}

We will consider a model where messengers transform only under one of the SM gauge groups $SU(3)\otimes SU(2)_L\otimes U(1)_Y$ and will choose (non-exotic) representations which are contained in $SU(5)$. In particular we choose the messenger representations~\cite{Delgado:2015bwa}
\begin{equation}
\Phi_{8}=(\mathbf{8},\mathbf{1})_0, \quad \Phi_3=(\mathbf{1},\mathbf{3})_0\quad \textrm{and} \quad \left[\Phi_1=(\mathbf{1},\mathbf{1})_1,\,\bar\Phi_1=(\mathbf{1},\mathbf{1})_{-1}\right]\ .
\end{equation}
According with GGM we will explore the more general case where the messengers have independent mass terms instead of getting all their mass from the spurion superfield. For simplicity, we also consider that the scalar component of $X$ does not acquire a VEV, thus $\langle X \rangle=\theta^2F$ and the superpotential couplings of messengers with the superfield $X$
\begin{equation}
W=\left({\lambda}_8 X+\mathcal M_8 \right)\Phi_{8}\Phi_{8}+\left({\lambda}_3X+\mathcal M_3\right)\Phi_{3}\Phi_{3}+\left({\lambda}_1 X+\mathcal M_1\right)\bar\Phi_{1}\Phi_{1}
\end{equation}
Moreover for simplicity we will consider a common messenger scale so that we will assume $\mathcal M_A\equiv \mathcal M$ ($A=8,3,1$).

Within this setup and with $\Lambda_8\equiv {\lambda}_8\Lambda$, $\Lambda_3\equiv {\lambda}_3\Lambda$ and $\Lambda_1\equiv {\lambda}_1\Lambda$ ($\Lambda\equiv F/\mathcal M$) the gaugino masses at the messenger scale are,
\begin{equation}
M_3=\frac{\alpha_3(\mathcal M)}{4\pi}3n_8\Lambda_8,\quad
M_2=\frac{\alpha_2(\mathcal M)}{4\pi}2n_3\Lambda_3 \quad
M_1=\frac{\alpha_1(\mathcal M)}{4\pi}\frac{6}{5}n_1\Lambda_1\, ,
\label{gauginos}
\end{equation}
where we are using $SU(5)$ normalization for the $U(1)$. The sfermion squared masses at the messenger scale are
\begin{equation}
m_{\tilde{f}}^2=2\left[C_3^f\left(\frac{\alpha_3(\mathcal M)}{4\pi}\right)^23n_8\Lambda_8^2+C_2^f\left(\frac{\alpha_2(\mathcal M)}{4\pi}\right)^22n_3\Lambda_3^2+ C_1^f\left(\frac{\alpha_1(\mathcal M)}{4\pi}\right)^2\frac{1}{2}\left(\frac{6}{5}\right)^2n_1\Lambda_1^2 \right] ,
\label{sfermions}
\end{equation}
where $n_8$, $n_3$ and $n_1$ are the of number of copies of each messenger respectively~\footnote{In the case of $n_1$, it is the number of pairs $(\Phi_1,\tilde{\Phi}_1)$ due to anomaly cancelation.}.

We can then write, at one loop, an RG invariant gaugino mass relation which will be different from the minimal case $M_1(\mathcal M)/\alpha_1(\mathcal M)=M_2(\mathcal M)/\alpha_2(\mathcal M)=M_3(\mathcal M)/\alpha_3(\mathcal M)$. In particular
\begin{equation}
\label{eq:gauginomass}
\frac{M_1(\mathcal M)}{\alpha_1(\mathcal M)}:\frac{M_2(\mathcal M)}{\alpha_2(\mathcal M)}:\frac{M_3(\mathcal M)}{\alpha_3(\mathcal M)}=\frac{6}{5}n_1\lambda_1:2n_3\lambda_3:3n_8\lambda_8\,.
\end{equation}
The free parameters are then $\lambda_A, n_A$ ($A=8,3,1$) and $\sqrt{F}$, $\mathcal M$.

In this scenario of low supersymmetry breaking the lightest supersymmetric particle (LSP) is the gravitino, with a mass
\begin{equation}
m_{3/2}\simeq \frac{F}{M_P}
\end{equation}
where $M_P=2.4\times 10^{18}$ GeV is the reduced Planck mass, it is therefore the dark matter candidate in the theory. The collider phenomenology depends on the mass of the next to lightest supersymmetric particle (NLSP). A quick glance at Eqs~(\ref{gauginos}) and (\ref{sfermions}) shows that the NLSP can be either the lightest neutralino $\tilde\chi_1^0$ or the right-handed stau $\tilde\tau_R$. We will use as bechmark scenarios those based on the ATLAS search for direct production of charginos, neutralinos and sleptons in final states with two leptons and missing transverse momentum~\cite{Aad:2014vma}.

\section{Benchmark scenarios in the SCTM}
Our main goal is to achieve light stop masses within the context of gauge mediation. Due to the strongest color contribution, if gluinos are heavier than stops they will raise the stop masses through the RG running, making their boundary condition at the messenger scale unimportant. In a gauge mediated context we can generally say that the heavier the gluino the heavier the stop. Therefore we will fix the gluino mass at the electroweak scale as low as possible consistently with the most stringent bounds released by the LHC data~\cite{Aad:2014mra} $M_3=1.5$ TeV. For the value of $\mathcal M=100$ TeV this will fix the supersymmetry breaking parameter $F$.

The NLSP will play an important role in the phenomenology of the model. In particular in each of the benchmark scenarios studied below, because of the low values of $\sqrt{F}$ the decay $NLSP\to \tilde G+...$ will be prompt, i.e.~it will decay inside the detector but with no displaced vertex, and the experimental signature will be an imbalance in the final state momenta and a pair of photons or charged leptons. Now from the ATLAS results there are two main scenarios:
\begin{description}
\item{\sc \underline{Scenario 1: Bino-like NLSP}}

\noindent $m_{\tilde \tau_R}>m_{\tilde\chi_1^0}$: in this case the experimental bounds are~\cite{Aad:2014vma}
\begin{equation}
m_{\tilde \tau_R}>m_{\tilde\chi_1^0}\gtrsim 100 \textrm{ GeV},\quad m_{\tilde \tau_L}>350  \textrm{ GeV}
\end{equation}
\item{\sc\underline{Scenario 2: Right-handed stau NLSP}}

\noindent $m_{\tilde \tau_R}<m_{\tilde\chi_1^0}$: in this case the experimental bounds are~\cite{Aad:2014vma}
\begin{equation}
m_{\tilde \tau_R}>250  \textrm{ GeV},\quad m_{\tilde \tau_L}>250  \textrm{ GeV}
\end{equation}
\end{description}

\subsection{Bino-like NLSP}
\label{scenario1}
This scenario is realized for the following values of the parameters
\begin{equation}
\label{eqn:benchmark1}
n_1=1,\, n_3=2, \,n_8=6\quad\textrm{and}\quad {\lambda}_1=0.9,\, {\lambda}_3=0.5, \,{\lambda}_8=0.1 \,. 
\end{equation}
The $SU(2)_L\otimes SU(2)_R$ invariant $\lambda$ of the superpotential will be fixed at the messenger scale such that the correct Higgs mass is reproduced, $\lambda(\mathcal M)=0.68$.
We also fix the superpotential parameter $\lambda_3(\mathcal{M})=0.35$, although it will have little effect on the low energy spectrum. The boundary conditions at the messenger scale of $\mu$ (and $\mu_\Delta$) are adjusted to make sure that the vacuum is close enough to the direction $\tan\theta_0=1$, and $\rho$ falls within the allowed $T$ parameter band, $T=0.01\pm 0.12$~\cite{Agashe:2014kda}. In this case we choose them to be
\begin{equation}
\label{eqn:benchmark3}
\mu(\mathcal{M})=\mu_\Delta(\mathcal{M})= 1.3\,\,\, \mathrm{TeV}
\end{equation}

\begin{figure}[htb]
\begin{center}
\vspace{0.5cm}
\includegraphics[width=0.49\linewidth]{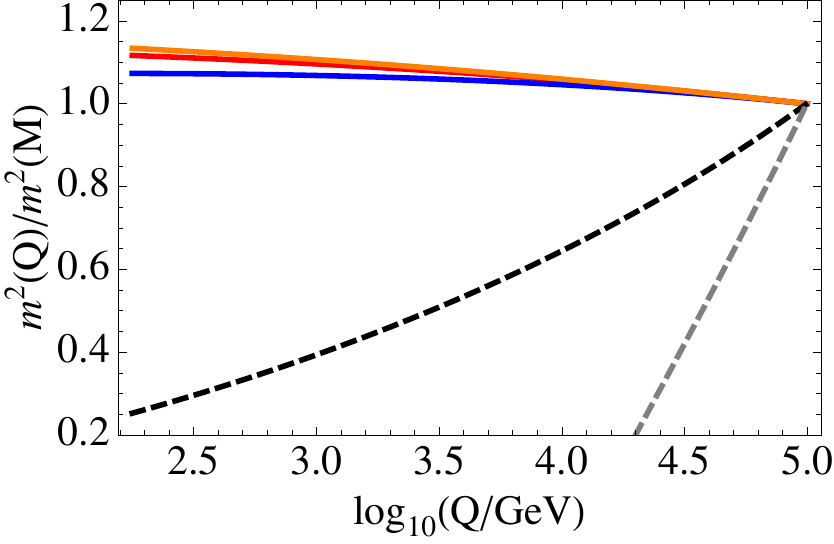}
\includegraphics[width=0.49\linewidth]{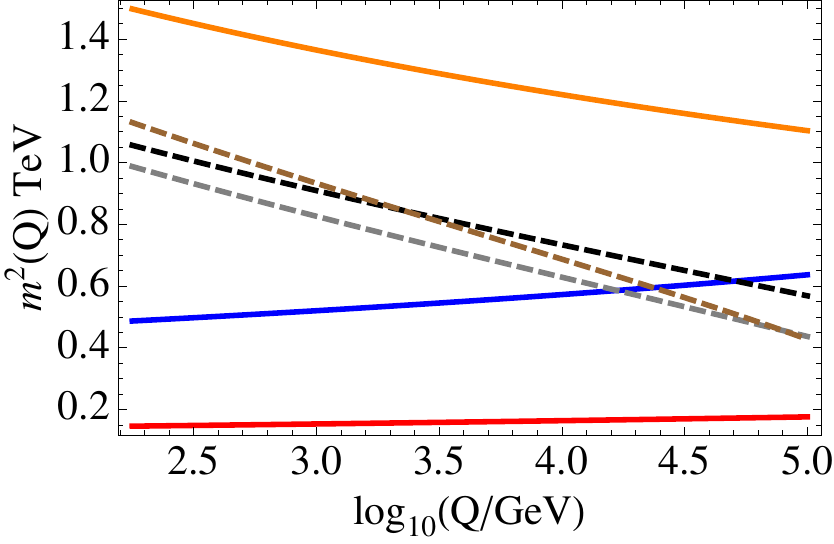}
\caption{\emph{Left panel: Running of $(m_{H_1}^2,m_{H_2}^2)$ (dashed lines) and $(m_{\Sigma_0}^2,m_{\Sigma_1}^2,m_{\Sigma_{-1}}^2)$ (solid lines), normalized to their values at the messenger scale for benchmark scenario 1. Right panel: Running of gaugino (solid: $M_3$ orange, $M_2$ blue and $M_1$ red) and squark (dashed: $m_{\tilde{Q}}$ black, $m_{\tilde{t}}$ gray and $m_\protect{\tilde{b}}$ brown) mass parameters for benchmark scenario 1.}}
\label{fig:running}
\end{center}
\end{figure}

Because of the strong effect of the top quark Yukawa coupling, the running differentiates the two soft doublet masses from each other much more than the three triplet ones among themselves. This behaviour which is explicitly shown in Fig.~\ref{fig:running} will result in a much bigger vacuum misalignment in the doublet sector. We are therefore left with a situation at the weak scale where $\tan{\beta}\neq 1$ and $(\tan{\theta_0},\tan{\theta_1}\sim 1)$ and so the loop contributions to the $\rho$ parameter coming from the doublet (MSSM) sector will be dominant.
%

In this scenario the NLSP will mainly decay to the gravitino through the following process $\chi_1^0 \rightarrow \gamma \tilde{G}$. If we know its mass and the supersymmetry breaking scale $\sqrt{F}$ we can calculate the average distance travelled in the LAB frame by an NLSP produced with energy $E$ before it decays. In this scenario $\sqrt{F}=94$ TeV and $m_{\chi_1^0}=143$ GeV, this translates in an average distance of flight $L_{\mathrm{Scenario 1}}^{NLSP}$well below the detector precision ($\sim 0.1$ cm) even if the particle is produced with very high energy and really boosted.

\begin{figure}[htb]
\begin{center}
\hspace{-0.2cm}\includegraphics[width=0.46\linewidth]{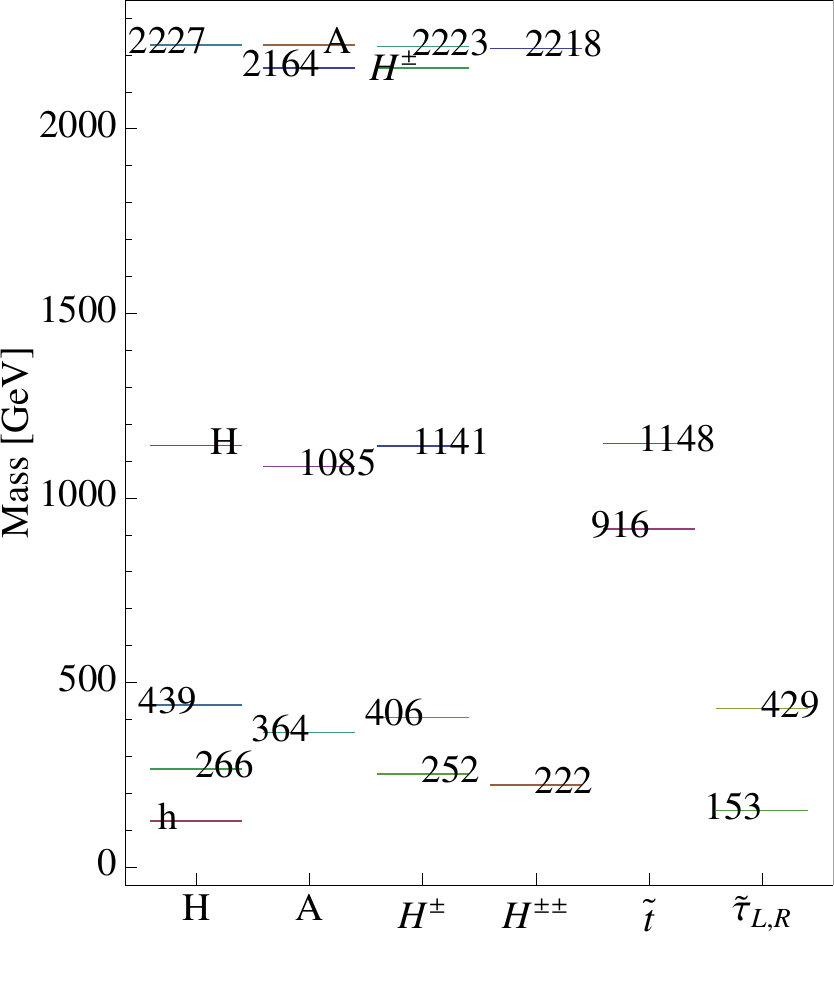}\hspace{5mm}
\includegraphics[width=0.46\linewidth]{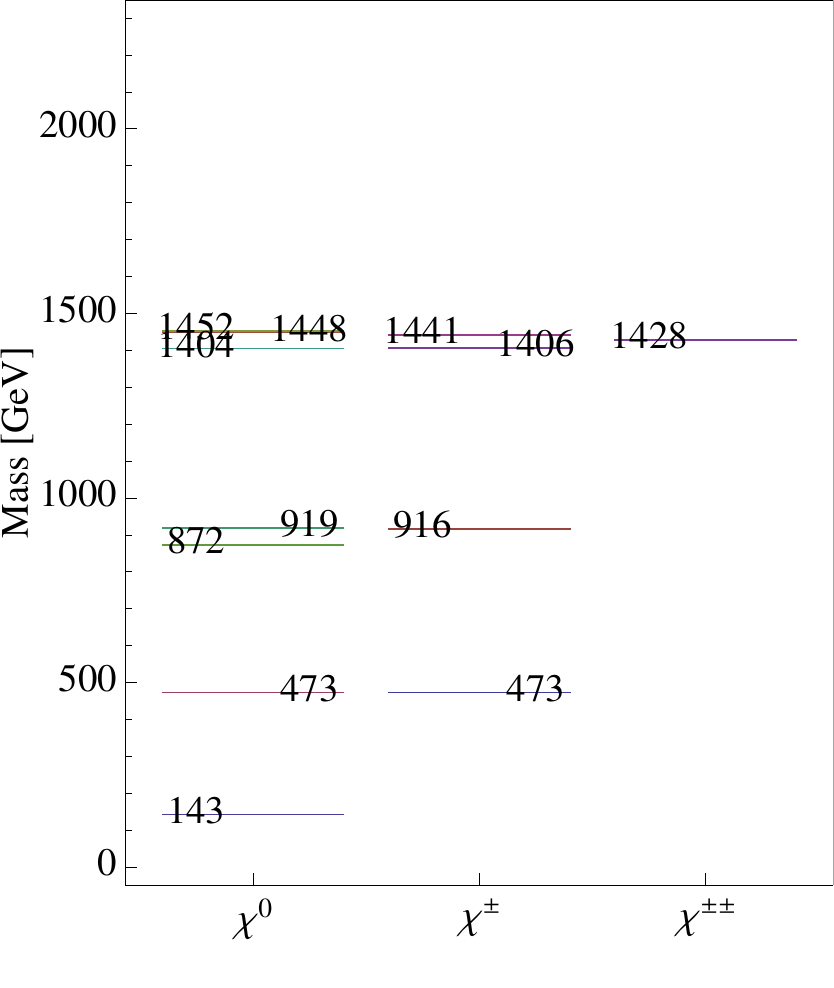}
\caption{\emph{Left panel: Scalar spectrum for scenario 1. MSSM-like scalars are quoted as so. Right panel: Fermion spectrum for scenario 1.
}}
\label{fig:spectrum}
\end{center}
\end{figure}

The fermionic spectrum satisfies the relation
\begin{equation}
\frac{M_1}{\alpha_1}:\frac{M_2}{\alpha_2}:\frac{M_3}{\alpha_3}=1.08:2:1.8\, \end{equation}
The lightest fermion is a Bino-like neutralino. The next neutralino and first chargino correspond to a Wino-like multiplet, since $M_2$ at the low scale is around $450-500$ GeV. In this scenario $\tilde\chi_2^0$ and the lightest chargino $\tilde\chi^\pm_1$ are (quasi) degenerate in mass. The ATLAS supersymmetric searches~\cite{Aad:2014vma} on $\tilde\chi^0_2\tilde\chi^\pm_1$ production followed by $W$ and $Z$ decays, combined with three-lepton searches, exclude a mass region for degenerate $\tilde\chi_2^0$ and  $\tilde\chi^\pm_1$ between 100 GeV and 410 GeV. These bounds are satisfied since the mass of $\tilde\chi_2^0$ and  $\tilde\chi^\pm_1$ is $\sim 473$ GeV. The heavier states are doublet-like Higgsinos and tripletinos. Fig.~\ref{fig:spectrum} shows the different mass values for Scenario 1.

The normalized couplings of the Higgs to vector bosons and fermions are defined as
\begin{equation}
r_{hXX}=\frac{g_{hXX}}{g_{hXX}^{SM}} \textrm{ with } X=V(W,Z),f(t,b,\tau) 
\end{equation}
From the values of $r_{hXX}$ one can also compute the predicted signal strength $\mu_{hXX}$ of the decay channel $h\to XX$, with $X=V,f,\gamma$:
\begin{equation}
\mu_{hXX}=\frac{\sigma(pp\to h)BR(h\to XX)}{[\sigma(pp\to h)BR(h\to XX)]_{SM}}
\end{equation}
In particular for the gluon-fusion (gF), the associated
production with heavy quarks ($htt$), the associated
production with vector bosons ($Vh$) and the vector boson
fusion (VBF) production processes, one can write
$\mu_{hXX}^{(gF)}=\mu_{hXX}^{(htt)}=
r_{hff}^2 r_{hXX}^2/\mathcal D$ and $\mu_{hXX}^{(VBF)}=\mu_{hXX}^{(Vh)}=r_{hVV}^2 r_{hXX}^2/\mathcal D$. Where 
$\mathcal D \simeq
0.74\, r_{hff}^2 + 0.26\, r_{hVV}^2$.
The Higgs couplings and signal strengths for scenario~1 are presented in Tab.~\ref{table1}.
\begin{table}[htb]
\begin{center}
\begin{tabular}{|c||c|c|c|c|c|}
\hline
Scenario 1 &${WW}$&${ZZ}$&${b\bar{b}}$&$t\bar{t}$&${\gamma\gamma}$\\ \hline \hline
$r_{hXX}$ &1.05 & 1.04 & 1.01& 1.01&1.22\\ \hline \hline
$\mu_{hXX}^{(gF)},\mu_{hXX}^{(htt)}$&1.07&1.05&1&0.99&1.45\\ \hline
$\mu_{hXX}^{(WF)},\mu_{hXX}^{(Wh)}$&1.16&1.14&1.08&1.07&1.58\\ \hline
$\mu_{hXX}^{(ZF)},\mu_{hXX}^{(Zh)}$&1.14&1.11&1.06&1.05&1.54\\ \hline
\end{tabular}
\end{center}
\caption{\em Higgs couplings and signal strengths for scenario 1. }
\label{table1}
\end{table}

\subsection{Right-handed stau NLSP}
\label{scenario2}

This scenario is realized for the following values of the parameters
\begin{equation}
\label{eqn:benchmark1}
n_1=10,\, n_3=6, \,n_8=5\quad\textrm{and}\quad {\lambda}_1=0.9,\, {\lambda}_3=0.5, \,{\lambda}_8=0.2 \,. 
\end{equation}
The $SU(2)_L\otimes SU(2)_R$ invariant $\lambda$ of the superpotential will be fixed at the messenger scale such that the correct Higgs mass is reproduced $\lambda(\mathcal M)=0.78$. We also fix the superpotential parameters $\lambda_3(\mathcal{M})=0.35$ and $\mu(\mathcal{M})=\mu_\Delta(\mathcal{M})= 1.5\,\,\, \mathrm{TeV}$. The $\tilde{\tau}$ will decay into the gravitino through $\tilde{\tau} \rightarrow \tau \tilde{G}$. In this case $\sqrt{F} =73$ TeV and $m_{\tilde{\tau}} =343$ GeV and one finds that $L_{\mathrm{Scenario 2}}^{NLSP} < L_{\mathrm{Scenario 1}}^{NLSP}$.

\begin{figure}[htb!]
\begin{center}
\hspace{-0.2cm}\includegraphics[width=0.46\linewidth]{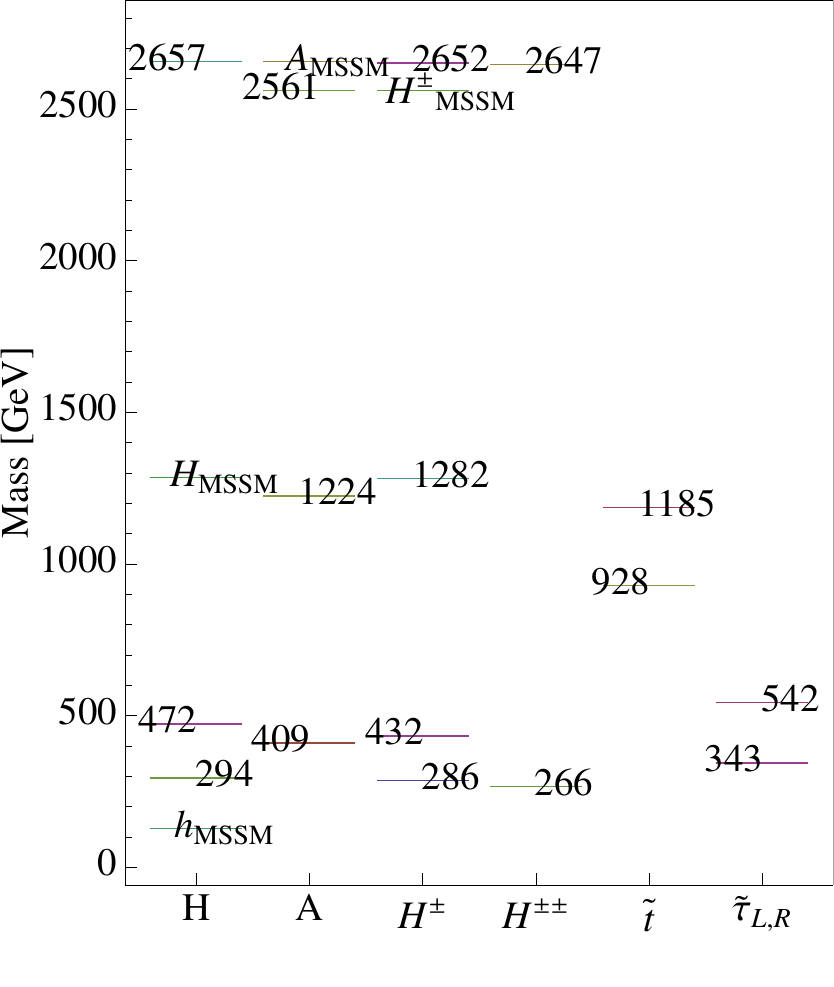}\hspace{5mm}
\includegraphics[width=0.46\linewidth]{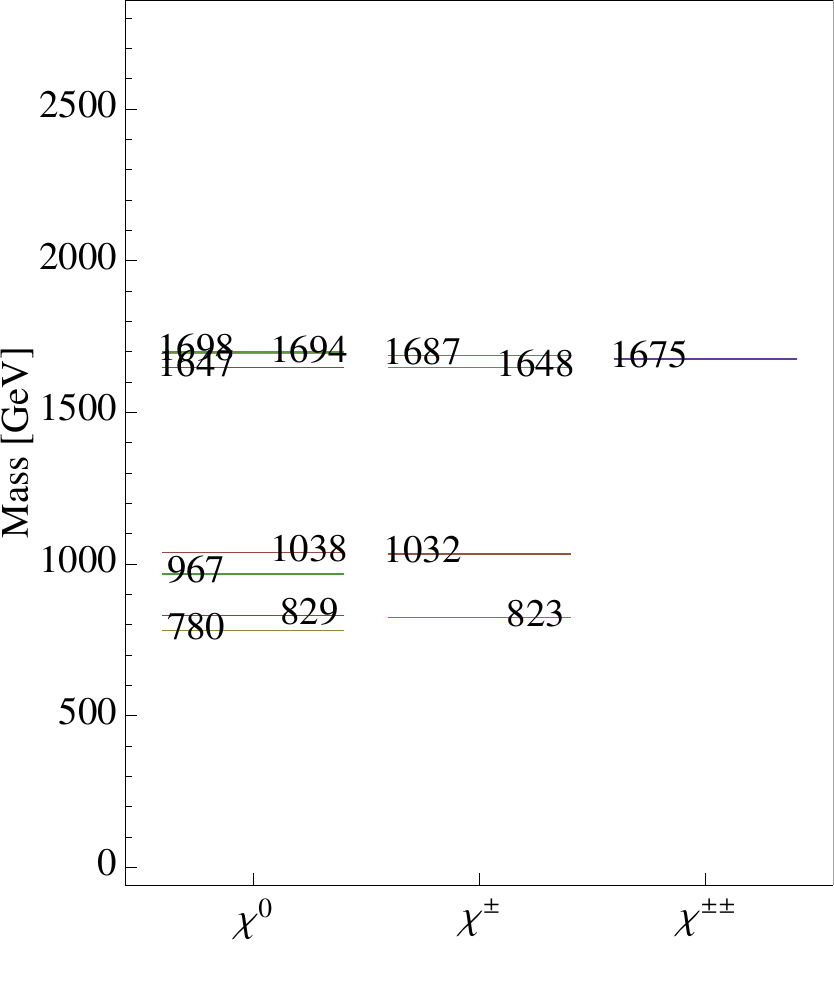}
\caption{\emph{Left panel: Scalar spectrum for scenario 2. MSSM-like scalars are quoted as so. Right panel: Fermion spectrum for scenario 2.
}}
\label{fig:spectrumbis}
\end{center}
\end{figure}

The fermion spectrum satisfies
\begin{equation}
\frac{M_1}{\alpha_1}:\frac{M_2}{\alpha_2}:\frac{M_3}{\alpha_3}=10.8:6:3
\end{equation}
and this different hierachy is explicit in Fig.~\ref{fig:spectrumbis}, with a fermion spectrum heavier than in the previous case, also satisfying all present experimental bounds. The Higgs couplings and signal strengths for scenario~2 are presented in Tab.~\ref{table2}.

\begin{table}[htb]
\begin{center}
\begin{tabular}{|c||c|c|c|c|c|}
\hline
Scenario 2 &${WW}$&${ZZ}$&${b\bar{b}}$&$t\bar{t}$&${\gamma\gamma}$\\ \hline \hline
$r_{hXX}$ &1.05 & 1.04 & 1.01& 1.0&1.17\\ \hline \hline
$\mu_{hXX}^{(gF)},\mu_{hXX}^{(htt)}$&1.07&1.06&0.99&0.95&1.35\\ \hline
$\mu_{hXX}^{(WF)},\mu_{hXX}^{(Wh)}$&1.16&1.15&1.08&1.05&1.46\\ \hline
$\mu_{hXX}^{(ZF)},\mu_{hXX}^{(Zh)}$&1.15&1.14&1.07&1.03&1.45\\ \hline
\end{tabular}
\end{center}
\caption{\em Higgs couplings and signal strengths for scenario 2. }
\label{table2}
\end{table}

\subsection{Final comments}
Both scenarios are in agreement with the ATLAS current measurements within the present uncertainties. However as the precision will increase, the measurements of Higgs properties will offer one of the most promising avenues to probe this model. The Higgs is a doublet-like state and therefore its couplings to vector bosons and fermions will not be greatly modified, since the rest of the doublet-like spectrum is heavy enough. However because custodial invariance is broken at the electroweak scale by the RG running it turns out that there is a corresponding breaking of universality as the parameter $\lambda_{WZ}=r_{WW}/r_{ZZ}$ departs from one. In particular as we can see from Tab.~\ref{table1}, $\lambda_{WZ}-1\simeq 1\%$ for the benchmark scenario 1 and $\lambda_{WZ}-1\simeq 3\%$ for the benchmark scenario 2. This breaking of universality was considered in Ref.~\cite{Garcia-Pepin:2014yfa} as one of the possible smoking guns of our model.

Loop induced couplings like $h\gamma\gamma$ can have large modifications. New charged triplet-like light scalar states like $H^\pm$ or $H^{\pm\pm}$ are present and will modify the coupling by circulating along the loop. The lighter these particles are, the greater their effect will be in $r_{h\gamma\gamma}$ and since the masses of triplet-like states scale with $v_\Delta$, $h\rightarrow \gamma\gamma$ will soon put bounds on $v_\Delta$.

\section{Smoking guns}

In our model there is an extended \textit{fermiophobic} triplet Higgs sector, absent from the usual supersymmetric extensions of the Standard Model, whose neutral components can acquire a sizeable VEV $v_\Delta$.  As a consequence there is a rich phenomenology by new singly ($H^\pm$) and doubly charged ($H^{\pm\pm}$) scalars which, if light can contribute sizeably in loops to $r_{\gamma\gamma}$. On top of that there are two main signatures very characteristic of the model which can be considered as smoking guns of it.
\begin{itemize}
\item
The couplings $H^\pm W^\mp Z$ and $H^{\pm\pm}W^\mp W^\mp$ are proportional to $v_\Delta$ and can thus provide unique signatures for models with extended Higgs sector contributing to the electroweak symmetry breaking mechanism
\begin{eqnarray}
&&H^\pm\to W^\pm Z,\quad  \textrm{ (absent for doublets in the MSSM or the 2HDM)}\nonumber\\
&&H^{\pm\pm} \to W^\pm W^\pm\ .
\end{eqnarray}

The composition of the lightest charged Higgs $H^\pm$ is shown in Fig.~\ref{fig:spectrumcharged}.
\begin{figure}[htb]
\begin{center}
\hspace{-0.2cm}\includegraphics[width=0.6\linewidth]{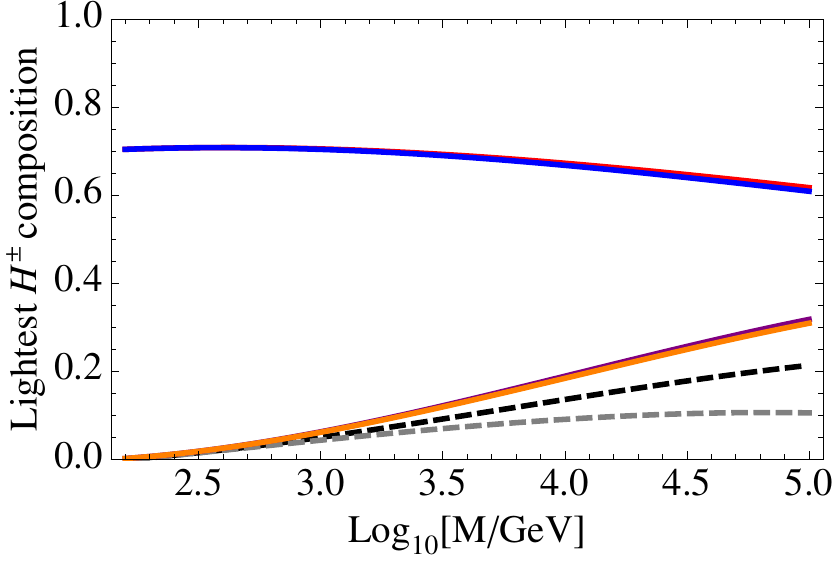}\hspace{5mm}
\caption{\emph{Plot that shows the composition of the lightest $H^\pm$ state for points computed with different values of the messenger mass scale, $\mathcal{M}$. Thick lines correspond to components coming from the triplet sector and dashed ones are components of the doublet sector. The lightest $H^\pm$ will be dominantly triplet like even including the custodial breaking caused by the running, this means that its couplings to gauge bosons will be weighted by the factor $\sim v_\Delta/v$. The other possible smoking gun of the model, the doubly charged state $H^{\pm\pm}$ will be totally triplet as there is no doubly charged component coming from the doublet sector.}}
\label{fig:spectrumcharged}
\end{center}
\end{figure}
\item
One can measure the amount of custodial breaking~\cite{Low:2010jp}
 by the departure of the universality parameter $\lambda_{WZ}\equiv r_{WW}/r_{ZZ}$ from one.
\end{itemize}

\section{Conclusion}
We will summarize the conclusions in the following items.
\begin{enumerate}
\item
Introducing triplets with hypercharges $Y=(0,\pm 1)$ permits to use GMSB with light ($\sim$ 1 TeV) stops, thus alleviating the little hierarchy problem of the MSSM.
\item
Custodial symmetry requires low-scale supersymmetry breaking. In our model we have considered $\sqrt{F}\simeq 70-90$ TeV.
\item
The typical pattern for the values of $M_a/\alpha_a$ is strongly spoiled. The spectrum is completely different from that of the typical MGM.

\item
The triplet VEVs  can contribute with a non negligible amount to the mechanism of electroweak breaking. A very interesting fact that will be explored by the LHC as well as the next generation colliders.
\item
The couplings $H^\pm W^\mp Z$ and $H^{\pm\pm}W^\mp W\mp$  (absent in the MSSM) are proportional to $v_\Delta$ and can thus provide unique signatures for models with extended Higgs sector contributing to the electroweak symmetry breaking mechanism.
\item
One can measure the amount of custodial breaking by the departure from one of the universality parameter $\lambda_{WZ}\equiv r_{WW}/r_{ZZ}$.
\item
It can provide, as type II seesaw models, a renormalizable neutrino Majorana mass term from the $\Delta L=2$ superpotential $W_\nu=h^{ij}_\nu L_i\Sigma_1 L_j$. This term would break explicitly custodial invariance but by a tiny amount due to the smallness of the Yukawa coupling $h_\nu$.
\end{enumerate}

\appendix

\section{Electroweak minimum}

We will here make a systematic study of the custodially broken minimum (induced by radiative corrections) at the electroweak scale starting from a theory at the scale $\mathcal M$ which is custodially symmetric.
For pedagogical reasons we will start with the case of the MSSM with custodial symmetry at $\mathcal M$.

\subsection{Warming up with the MSSM}
In the MSSM the Higgs potential is a function of two fields $V=V(h_2,h_1)$ the real parts of the neutral components of the Higgs doublets $(H_2,H_1)$. These two degrees of freedom will make up the $CP$-even MSSM mass eigenstates $(h,H)$. At the minimum the VEVs are defined as $v_2=v\sin\beta$ and $v_1=v\cos\beta$. The equations of minimum $\partial V/\partial H_{2,1}^0=0$ provide the equations
\begin{eqnarray}
&&m_2^2-m_3^2\cot\beta-\frac{m_Z^2}{2}\cos 2\beta=0\label{1MSSM}\\
&&m_1^2-m_3^2\tan\beta+\frac{m_Z^2}{2}\cos 2\beta=0\label{2MSSM}
 \end{eqnarray}
 where we are defining $m_{1,2}^2\equiv m_{H_1,H_2}^2+|\mu|^2$. Now the linear combinations Eq.~(\ref{1MSSM})~$\pm$~Eq.~(\ref{2MSSM}) lead respectively to

 \begin{eqnarray}
 m_3^2 & =&\frac{\tan\beta}{\tan^2\beta+1}(m_1^2+m_2^2)
 \label{1p2MSSM}\\
 \tan^2\beta&=&\frac{m_1^2+\frac{1}{2}m_Z^2}{m_2^2+\frac{1}{2}m_Z^2}
 \label{1m2MSSM}
 \end{eqnarray}
where we are assuming $m_1^2\geq m_2^2$.

Now we can first assume that the Higgs sector has custodial symmetry and that therefore $m_1=m_2$. In this case we see that Eq.~(\ref{1m2MSSM}) is identically satisfied for $\tan\beta=1$, which is precisely the custodial symmetric minimum, while Eq.~(\ref{1p2MSSM}) yields $m_3^2=(m_1^2+m^2_2)/2$ which is the condition for electroweak symmetry breaking (EWSB). 
Second, we will assume that the theory is custodially symmetric at the scale $\mathcal M$ where supersymmetry breaking is communicated to the observable sector. In this case, as there are couplings which do not respect the custodial symmetry (in particular $g_1$ and $h_t$), even if $m_1=m_2$ at $Q=\mathcal M$ at the EW scale, the latter equality will not hold. In this case at the EW scale a value $\tan\beta\neq 1$ will be generated 
%
%
and the value of $m_3^2$ will then be correspondingly obtained from Eq.~(\ref{1p2MSSM}).

\subsection{The Custodial Supersymmetric Triplet Model}
We will now apply the previous procedure to the case of the Supersymmetric Triplet model with custodial symmetry. The Higgs sector is custodially invariant at the scale $\mathcal M$ but the RGE running will spoil the custodial symmetry mainly because the couplings $(g_1,h_t)$ break it. So in principle (as the MSSM example above) we should write the most general potential for this theory. This can be done from the superpotential for the neutral components of the Higgs doublets $(H_1,H_2)$ and triplets $(\Sigma_1,\Sigma_0,\Sigma_{-1})$ with superpotential given in Eq.~(\ref{superpotencial}).
%
%
The potential is then $V=V_F+V_D+V_{\mathrm{soft}}$ where $V_F$ is the supersymmetric potential obtained from the superpotential (\ref{superpotencial}), 
\begin{equation}
V_D=\frac{g_2^2+g_1^2}{8}(|H_1|^2-|H_2|^2+2 |\chi|^2-2|\psi|^2)^2
\label{potencialD}
\end{equation}
and
\begin{eqnarray}
V_{\mathrm{soft}}&=&m_{H_1}^2H_1^\dagger H_1+m_{H_2}^2H_2^\dagger H_2+m_{\Sigma_0}^2\Sigma_0^\dagger \Sigma_0+m_{\Sigma_1}^2\Sigma_1^\dagger \Sigma_1+m_{\Sigma_{-1}}^2\Sigma_{-1}^\dagger \Sigma_{-1} -  m_3^2 H_1\cdot H_2 \nonumber\\ 
&+&\left\{\frac{B_{a}}{2}\textrm{tr}\Sigma_0^2+B_{b}\textrm{tr}\Sigma_1\Sigma_{-1}-A_{a}H_1\cdot\Sigma_1H_1+A_{b}H_2\cdot\Sigma_{-2}H_2\right. \nonumber \\ 
&+&\left.\sqrt{2}A_{c}H_1\cdot\Sigma_0H_2+\sqrt{2}A_{\lambda_3}\textrm{tr}\,\Sigma_1\Sigma_0\Sigma_{-1}+ a_t\,\tilde Q_L\cdot H_2 \tilde t_R+ a_b\,\tilde Q_L\cdot H_1\tilde{b}_R+h.c.\right\}
\label{potencialsoft}
\end{eqnarray}

The custodial invariance translates into the following boundary conditions at $Q=\mathcal M$
\begin{eqnarray}
\lambda_a&=&\lambda_b=\lambda_c\equiv \lambda\nonumber\\
\mu_a&=&\mu_b\equiv \mu_\Delta\nonumber\\
m_{H_2}&=&m_{H_1}\equiv m_H\nonumber\\
m_{\Sigma_0}&=&m_{\Sigma_1}=m_{\Sigma_{-1}}\equiv m_\Delta\nonumber\\
A_a&=&A_b=A_c\equiv A_\lambda\nonumber\\
B_a&=&B_b\equiv B_\Delta
\end{eqnarray}
The EoM are the solutions to the equations $\partial V/\partial H_1=\partial V/\partial H_2=\partial V/\partial \psi=\partial V/\partial \phi=\partial V/\partial \chi=0$ which are satisfied at the (real part of the) field VEVs $(h_1,h_2,\psi,\phi,\chi)=(v_1,v_2,v_\psi,v_\phi,v_\chi)$.

The equation $(1/H_1)\partial V/\partial H_1+(1/H_2)\partial V/\partial H_2=0$ allows to trade the parameter $m_3^2$ by the other supersymmetric parameters, as
\begin{eqnarray}
&&\frac{1}{\sin 2\beta}\left[m_3^2-A_c v_\phi -\lambda_c(\lambda_3 v_\chi v_\Psi+\lambda_c v_1 v_2+\mu_a v_\phi) -2(\lambda_c v_\phi-\mu)(\lambda_a v_\psi+\lambda_b v_\chi)\right]\nonumber\\
&=&\frac{1}{2}(m_{H_u}^2+m_{H_d}^2)
+A_a v_\psi+ A_b v_\chi+(\lambda_3 v_\phi+\mu_b)(\lambda_b v_\psi+\lambda_a v_\chi)\nonumber\\
&+&\lambda_a^2 v_1^2+\lambda_b^2 v_2^2+(\lambda_c v_\phi-\mu)^2
+2\left[\lambda_a^2 v_\psi^2+\lambda_b^2 v_\chi^2\right]
\label{m32}
\end{eqnarray}
where $\tan\beta=v_2/v_1$.
The value of $m_3^2$ from Eq.~(\ref{m32}) is now replaced into the equation $(1/H_1)\partial V/\partial H_1-(1/H_2)\partial V/\partial H_2=0$ which then becomes
\begin{eqnarray}
&&\frac{g_1^2+g_2^2}{2}(v_2^2-v_1^2+2v_\psi^2-2 v_\chi^2)=2\cos 2\beta(\lambda_c v_\phi-\mu)^2
\nonumber\\
&+&2\cos^2\beta\left\{m_{H_d}^2+2A_a v_\psi+2\lambda_a(\lambda_3 v_\phi+\mu_b)v_\chi+2\lambda_a^2 (v_1^2+2 v_\psi^2)
 \right\}\nonumber\\
&-&2\sin^2\beta\left\{m_{H_u}^2+2A_b v_\chi+2\lambda_b(\lambda_3 v_\phi+\mu_b)v_\psi
+2\lambda_b^2 (v_2^2+2 v_\chi^2)
  \right\}
\label{eq1}
\end{eqnarray}
This equation is identically satisfied in the custodial limit.

Likewise, equation $(1/\psi) \partial V/\partial \psi+(1/\chi)\partial V/\partial \chi$ yields the parameter $B_b$ as a function of the other supersymmetric parameters as
\begin{eqnarray}
&&\frac{v_\psi^2+v_\chi^2}{v_\psi v_\chi}
\left[-B_b-A_3v_\phi-\lambda_3(\lambda_c v_1 v_2+\mu_a v_\phi+\lambda_3 v_\psi v_\chi)\right]=\nonumber\\
&&m_{\Sigma_1}^2+m_{\Sigma_{-1}}^2+(\lambda_3 v_\phi+\mu_b)\left(\frac{\lambda_a v_1^2}{v_\chi}+\frac{\lambda_b v_2^2}{v_\psi}  \right)+2(\lambda_3 v_\phi+\mu_b)^2+4(\lambda_a^2 v_1^2+\lambda_b^2 v_2^2)\nonumber\\
&+&\frac{A_a v_1^2+2\lambda_a v_ 1 v_ 2(\lambda_c v_\phi -\mu)}{v_\psi}
+\frac{A_b v_2^2+2\lambda_b v_ 1 v_ 2(\lambda_c v_\phi -\mu)}{v_\chi}
\label{Bb}
\end{eqnarray}
%
%
%
The value of $B_b$ is then replaced into equation $(1/\psi) \partial V/\partial \psi-(1/\chi)\partial V/\partial \chi$ which then becomes
\begin{eqnarray}
&&(g_1^2+g_2^2)(v_2^2-v_1^2+2 v_\psi^2-2 v_\chi^2)=-2\,
\frac{v_\psi^2-v_\chi^2}{v_\psi^2+v_\chi^2}\,
(\lambda_3 v_\phi+\mu_b)^2\nonumber\\
&-&2\,
\frac{v_\psi^2}{v_\psi^2+v_\chi^2}\,
\left\{m_{\Sigma_1}^2+4\lambda_a^2 v_1^2+(\lambda_3 v_\phi+\mu_b)\frac{\lambda_b v_2^2}{v_\psi}+\frac{A_a v_1^2+2\lambda_a v_1 v_ 2(\lambda_c v_\phi-\mu)}{v_\psi}\right\}\nonumber\\
&+&2\,
\frac{v_\chi^2}{v_\psi^2+v_\chi^2}\,
\left\{m_{\Sigma_{-1}}^2+4\lambda_b^2 v_2^2+(\lambda_3 v_\phi+\mu_b)\frac{\lambda_a v_1^2}{v_\chi}+\frac{A_b v_2^2+2\lambda_b v_1 v_ 2(\lambda_c v_\phi-\mu)}{v_\chi}\right\}
\label{eq2}
\end{eqnarray}
Again this equation is identically satisfied in the custodial limit.
Finally the value of $B_a$ is obtained from the equation $\partial V/\partial \phi=0$ as
\begin{eqnarray}
&-&\left[B_a+\mu_a^2+m_{\Sigma_0}^2\right]v_\phi=(A_c+\lambda_c \mu_a)v_1 v_2+(A_3+\lambda_3 \mu_a)v_\psi v_\chi\nonumber\\
&+&\lambda_c v_1(2\lambda_b v_ 2 v_\chi+\lambda_c v_ 1 v_\phi-v_ 1\mu)+\lambda_c v_2(2\lambda_a v_ 1 v_\psi+\lambda_c v_ 2 v_\phi-v_ 2\mu)\nonumber\\
&+& \lambda_3 v_\chi\left[\lambda_a v_1^2+v_\chi(\lambda_3 v_\phi+\mu_ b)  \right]+
 \lambda_3 v_\psi\left[\lambda_b v_2^2+v_\psi(\lambda_3 v_\phi+\mu_ b)  \right]
\label{Ba}
\end{eqnarray}
If we define $B_{\mp}=B_a\mp B_b$ then the EoM for $B_-$ is
\begin{eqnarray}
B_-&=& A_3\left(v_\phi-\frac{v_\psi v_\chi}{v_\phi} \right)+\frac{1}{v_\psi^2+v_\chi^2}\left[
v_\psi v_\chi(m_{\Sigma_1}^2+m_{\Sigma_{-1}}^2)-(v_\psi^2+v_\chi^2)m_{\Sigma_0}^2\right]\nonumber\\
&+&\lambda_3 \left[ \lambda_c v_1 v_2+\lambda_a v_1^2\left( \frac{v_\phi v_\psi}{v_\psi^2+v_\chi^2}-\frac{v_\chi}{v_\phi}\right)+\lambda_b v_2^2\left(\frac{v_\phi v_\chi}{v_\psi^2+v_\chi^2}-\frac{v_\psi}{v_\phi}\right)\right]\nonumber\\
&+&\lambda_3^2\left[ v_\psi v_\chi \frac{2 v_\phi^2+v_\psi^2+v_\chi^2}{v_\psi^2+v_\chi^2}-(v_\psi^2+v_\chi^2) \right]+\mu_b^2\, \frac{2 v_\psi v_\chi}{v_\psi^2+v_\chi^2}-\mu_a^2
\nonumber\\
&+& \lambda_3 v_\phi\left[ \mu_a\left(1-\frac{v_\psi v_\chi}{v_\phi^2}\right)+\mu_b\left(\frac{4v_\psi v_\chi}{v_\psi^2+v_\chi^2}-\frac{v_\psi^2+v_\chi^2}{v_\phi^2}\right) 
\right]\nonumber\\
&+&\frac{A_a v_1^2 v_\psi+A_b v_2^2 v_\chi}{v_\psi^2+v_\chi^2}-A_c\frac{v_1 v_2}{v_\phi}
+\mu_b \frac{\lambda_a v_1^2 v_\psi+\lambda_b v_2^2 v_\chi}{v_\psi^2+v_\chi^2}-\mu_a\frac{\lambda_c v_1 v_2}{v_\phi}\nonumber\\
&+&(\lambda_c v_\phi-\mu)\left[2 v_ 1 v_ 2\,\frac{\lambda_a v_\chi+\lambda_b v_\psi}{v_\psi^2+v_\chi^2}-\lambda_c\frac{v_1^2+v_2^2}{v_\phi}\right]\nonumber\\
&+&4(\lambda_a^2 v_1^2+\lambda_b^2 v_2^2)\,\frac{v_\psi v_\chi}{v_\psi^2+v_\chi^2}-2\lambda_c v_ 1 v_ 2\, \frac{\lambda_a v_\psi+\lambda_b v_\chi}{v_\chi}
\label{eq3}
\end{eqnarray}
which is also identically satisfied in the custodial limit. In fact we have written the different lines of Eq.~(\ref{eq3}) in such a way that they cancel independently  in the custodial limit. Finally the parameters $m_3^2$, and $B_+$ are given by Eqs.~(\ref{m32}), and (\ref{Bb}) and (\ref{Ba}), respectively. Eqs.~(\ref{eq1}), (\ref{eq2}), and (\ref{eq3}), which are identically satisfied in the custodial limit, will be used to compute the departure from the custodial symmetry triggered by the RGE running.

As Eqs.~(\ref{eq1}), (\ref{eq2}), and (\ref{eq3}) do not depend on the parameters $m_3^2$ and $B_+$, we will use them to compute the departure of the vacuum solution with respect to the custodial configuration by considering the general field configuration
\begin{eqnarray}
\tan\beta&=&\frac{v_2}{v_1},\quad v_1(\beta)=\sqrt{2}\cos\beta v_H,\quad v_2(\beta)=\sqrt{2}\sin\beta v_H\nonumber\\
\tan\theta_1&=&\frac{v_\chi}{v_\psi},\quad \tan\theta_0=\frac{\sqrt{2}v_\phi}{\sqrt{v_\psi^2+v_\chi^2}}\nonumber\\
v_\psi&=&2\cos\theta_1\cos\theta_0 v_\Delta,\quad v_\chi=2\sin\theta_1\cos\theta_0 v_\Delta,\quad v_\phi=\sqrt{2} \sin\theta_0 v_\Delta
\label{vacio}
\end{eqnarray}
where we have introduced two Euler angles $\theta_0$ and $\theta_1$ characterizing  the triplet VEV direction. Notice that
$v^2=v_1^2+v_2^2+2(2v_\phi^2+v_\psi^2+v_\chi^2)=2 v_H^2+8v_\Delta^2$ (where $v=174$ GeV) for any value of $\tan\beta$ and $\tan\theta_{1,0}$ so that one can trade the parameter $v_H$ by $v_\Delta$. In fact from the configuration in Eq.~(\ref{vacio}) the breaking of custodial symmetry (and the value of the $T$ parameter) is measured by $(\tan^2\theta_0-1)$ as
\begin{equation}
\alpha T=\frac{2v_\phi^2-(v_\psi^2+v_\chi^2)}{\frac{1}{2}(v_1^2+v_2^2)+2(v_\psi^2+v_\chi^2)}=
-4\frac{\cos 2\theta_0\, v_\Delta^2}{v_H^2+8\cos^2\theta_0 v_\Delta^2}
\label{T}
\end{equation}

Using the field configuration of Eq.~(\ref{vacio}) we can write an explicit solution to Eq.~(\ref{eq1}) as
\begin{equation}
\tan^2\beta=\frac{P_a-P_b+\sqrt{(P_a-P_b)^2+4 Q_aQ_b}}{2Q_b}
\label{beta}
\end{equation}
where $P_{a,b}$ and $Q_{a,b}$ are given by
\begin{eqnarray}
P_a&=&m_{H_1}^2+2 A_a v_\psi+2 \lambda_a v_\chi(\lambda_3 v_\phi+\mu_b)+4\lambda_a^2 v_\psi^2+(\lambda_c v_\phi-\mu)^2+\frac{g_1^2+g_2^2}{2}(v_H^2 +v_\chi^2-v_\psi^2)\nonumber\\
P_b&=&m_{H_2}^2+2 A_b v_\chi+2 \lambda_b v_\psi(\lambda_3 v_\phi+\mu_b)+4\lambda_b^2 v_\chi^2+(\lambda_c v_\phi-\mu)^2+\frac{g_1^2+g_2^2}{2}(v_H^2 +v_\psi^2-v_\chi^2)\nonumber\\
Q_{a}&=&P_{a}+4v_H^2\lambda^2_{a},\quad Q_{b}=P_{b}+4v_H^2\lambda^2_{b}
\end{eqnarray}
and where $v_{\phi,\psi}$ depend on $\theta_{0,1}$ through Eq.~(\ref{vacio}). In the custodial limit Eq.~(\ref{beta}) yields $\tan\beta=1$. Notice that Eq.~(\ref{beta}) is a straightforward generalization of the similar one for the MSSM, Eq.~(\ref{1m2MSSM}).

Now on general grounds Eqs.~(\ref{eq1}), (\ref{eq2}), and (\ref{eq3}) should be solved numerically, after running the RGE, to get the correct values of $\tan\beta$, $\tan\theta_0$ and $\tan\theta_1$. Eq.~(\ref{m32}) will determine the value of $m_3^2(Q_{EW})$ and the equation for $B_+$ [a linear combination of Eqs.~(\ref{Ba}) and (\ref{Bb})] will fix the custodial value $B_\Delta$ at the high scale $\mathcal M$.

\section*{Acknowledgments}\noindent
We would like to thank Roberto Vega-Morales for useful discussions on the nature of the $H^\mp W^\pm Z$ coupling. This research was supported in part  by the National Science Foundation under Grant No.~PHY-1215979, by the Spanish Consolider-Ingenio 2010 Programme CPAN
(CSD2007-00042), by CICYT-FEDER-FPA2011-25948, by the Severo Ochoa
excellence program of MINECO under the grant SO-2012-0234 and by
Secretaria d'Universitats i Recerca del Departament d'Economia i
Coneixement de la Generalitat de Catalunya under Grant 2014 SGR 1450.

\end{document}